\documentstyle[twocolumn,aps]{revtex}

\begin{document}
\draft

\twocolumn[\hsize\textwidth\columnwidth\hsize\csname @twocolumnfalse\endcsname

\title{Pair breaking by impurities in the two-dimensional $t-J$ model}

\author{J. Riera$^{1}$, S. Koval$^{1}$, D. Poilblanc$^{2}$ and F.
Pantigny$^{2}$}
\address{
$^{1}$Instituto de F\'{\i}sica Rosario, Consejo Nacional de
Investigaciones Cient\'{\i}ficas y T\'ecnicas
y Departamento de F\'{\i}sica, \\
Universidad Nacional de Rosario,
Blvd. 27 de Febrero 210bis,  2000-Rosario, Argentina\\
$^{2}$Laboratoire de Physique Quantique, Universit\'e Paul Sabatier,
31062 Toulouse, France
}

\date{October 95}
\maketitle

\begin{abstract}
Pair breaking mechanisms by impurities are investigated in the
two-dimensional $t-J$ model by exact diagonalization techniques.
Analysis of binding energies, pairing correlations, dynamical
spin and pair susceptibilities shows that non-magnetic
impurities are more effective in suppressing pairing than magnetic
ones in agreement with experimental studies of Zn- and Ni-
substituted High-Tc superconductors.
\end{abstract}

\pacs{ PACS Numbers: 74.20.Mn, 74.62.Dh, 71.55.-i, 75.40.Mg}
\vskip2pc]
\narrowtext

Many recent experiments of hole-doped cuprate superconductors
showing a large anisotropy consistent with gap nodes on the
Fermi surface\cite{expdwave} have been interpreted in terms of
an unconventional $d$-wave pairing state.
This possibility of $d$-wave superconductivity has in turn renewed
the interest in studying the effects of impurities on the
superconducting state.
It is well-known that magnetic impurities are strong pair breakers
for singlet superconductors while nonmagnetic
impurities have a pair breaking effect only for higher orbital
momentum states such as a $d$-wave pairing state.\cite{ueda}
In fact, in several recent studies it was suggested the possibility
of using the response of these materials to different types of
impurities to distinguish between an unconventional $d$-wave pairing
and alternative scenarios such as a more conventional anisotropic
$s$-wave pairing.\cite{borkowski,balatsky,fehrenorm}
Experimental studies have shown that divalent Zn and Ni ions
go to the planar Cu(2) sites with the Zn impurity having a $S=0$
configuration and the Ni impurity a $S=1$ configuration.\cite{xiao}
The important experimental fact is that the nominally nonmagnetic
impurity Zn is more effective than the magnetic impurity Ni in
destroying superconductivity.\cite{xiao,achkir,ulm}
In conventional superconductors, with
the usual $s$-wave superconductivity, a magnetic impurity is most
destructive due to the magnetic pair-breaking effect.\cite{abrikosov}
Although the observed behavior could be characteristic of a $d$-wave
pairing, it has been also suggested that a Zn impurity may not simply
behave as a vacancy because it could induce a magnetic moment in the
$\rm CuO_2$ planes due to the strong correlations
present.\cite{borkowski,xiao,mahajan}
However, other experimental studies\cite{walstedt} have indicated
that the estimated impurity moment- carrier exchange in Zn doped
YBa$_2$Cu$_3$O$_7$ is too small to account for the T$c$ suppression
by a magnetic pair-breaking mechanism.

{}From a theoretical point of view it is of extreme importance to
explain at least qualitatively the effects of both magnetic and
non-magnetic impurities on the superconducting properties of the
cuprates as seen experimentally.
These effects have not been clearly explained by existing analytical
macroscopical theories.
\cite{borkowski,balatsky,fehrenorm,hirsgold,hotta,kim,sunmaki}
An alternative approach is to analyze them in the context of
microscopic models using numerical techniques.
This program has been recently initiated by Poilblanc, Scalapino
and Hanke\cite{didier1,didier2} by studying the two-dimensional
(2D) $t - J$ model in the presence of a single impurity.
The two-dimensional $t - J$ model has been extensively
studied in the context of high-T$_c$ superconductivity
since it contains the essential low-energy physics of the CuO$_2$
planes present in the cuprates.\cite{dagorev}
The $S=0$ Zn impurity was modelled by an inert site,\cite{didier1}
while the $S=1$ Ni impurity was approximated by a static spin-1/2
(see Hamiltonian (\ref{hamtji}) below).\cite{didier2}
The main results obtained by these authors is that a hole binds to
the impurity with different spatial symmetries.

The purpose of this Letter is to study the effect of magnetic and
nonmagnetic impurities on several quantities related to pairing of
holes in the 2D $t-J$ model near half-filling. This study at zero
temperature will be performed using exact diagonalization techniques
on finite clusters. Our starting point is the fact that
numerical studies of the pure $t-J$ model on the square lattice have
shown that a two-hole bound state is formed for $J > J_c \sim 0.3\ t$
in the bulk limit and with d$_{x^2-y^2}$ internal
symmetry.\cite{binding,pairdyn}. These studies have also
given strong unbiased indications of d$_{x^2-y^2}$ superconductivity
at quarter filling ($n = 0.5$) in the vicinity of phase
separation,\cite{dagrieprl,ohta} although they are less conclusive
near half-filling. Our objective is also to ellucidate which is the
pair-breaking mechanism specially for the case of nonmagnetic
impurities.

The Hamiltonian of the $t-J$ model in the presence of impurities
is:\cite{didier2}
\begin{eqnarray}
\nonumber
H = &-& t \sum_{<i,j>,i,j\neq i_0;\sigma}
    ( \tilde{c}_{j \sigma}^\dagger \tilde{c}_{i \sigma}
    + \tilde{c}_{i \sigma}^\dagger \tilde{c}_{j \sigma} ) \\
\nonumber
   &+& J \sum_{<i,j>,i,j\neq i_0}
   ( {\bf S}_{i} \cdot {\bf S}_{j}
   - \textstyle{1 \over 4} n_{i} n_{j} )    \\
   &+& J^\prime \sum_{<i_0,j>}
   ( {\bf S}_{i_0} \cdot {\bf S}_{j}
   - \textstyle{1 \over 4} n_{i_0} n_{j} )
\label{hamtji}
\end{eqnarray}
\noindent
where $\tilde{c}_{i \sigma}^\dagger$ is an electron creation operator
at site $i$ with spin $\sigma$ with the constraint of no double
occupancy, $n_i = n_{i,\uparrow} + n_{i,\downarrow}$ is the electron
number operator. The impurities are located at the sites $\{i_0 \}$.
Here $t$ is the hopping parameter, which we choose as the scale of
energies, $J$ is the AF exchange interaction and
$J^\prime$ is the magnetic coupling between the impurity spin and the
electrons in nearest neighbor (NN) sites. The special case
$J^\prime =0$ corresponds to an inert site describing a Zn
impurity.\cite{wang}
We adopt periodic boundary conditions throughout.

As a preliminary study, it is interesting to investigate the effect
of the magnitude of the impurity spin $S_{i_0}$. Following previous
work, we define the hole-impurity binding energy\cite{didier1},
\begin{eqnarray}
E_{B,1} = e(1h,1i) - e(0h,1i) - e(1h,0i) ,
\end{eqnarray}
\noindent
where $e(Nh,Mi) = E(Nh,Mi) - E(0h,0i)$, $E(Nh,Mi)$ is the ground
state energy of the system with $N$ holes and $M$ impurities.
For the case of a spin 1 impurity embedded in small clusters,
bound states are found (i.e. $E_{B,1}<0$) in all symmetry channels
provided $\mid J^\prime/J\mid$ does not exceed small critical values.
This is qualitatively similar to the spin 1/2 impurity\cite{didier2}
as clearly seen in Fig. \ref{fig1} showing a comparison of
$E_{B,1}$ calculated on a 20-site cluster in the d-wave channel for
spin 1/2 and spin 1 impurities. Note that the coupling $J^\prime$ is
more effective to destroy the bound-state for larger impurity spin.
This can be simply understood from a simple argument relating the
binding energy with
the number of broken bonds in the AF background\cite{didier1}
(this also applies to pairing of holes \cite{dagorev}).
Indeed, while the average magnetic energy per bond for the bonds
not connected to the impurity\cite{bulut} depends weakly on
$J^\prime$, the magnetic energy for the bonds connected to the
impurity essentially scales like $J^\prime S_{i_0}$ so that the
effective short range attractive potential for holes weakens for
increasing $J^\prime$ or $S_{i_0}$.
In the following, we shall restrict ourself to the lowest energy
sector.
Moreover, since from Fig. \ref{fig1} the results for $S=1/2$ and
$S=1$ are very similar for the region of interest $J^\prime > 0$,
we shall consider only the case of spin-1/2 impurities.

In order to determine the binding of holes in pairs
in the vicinity of impurities let us start examining the following
combinations of ground state energies:
\begin{eqnarray}
\nonumber
&E_{B,2}& = e(2h,0i) - 2 e(1h,0i)    \\
&E^\prime_{B,2}& = e(2h,0i) + e(0h,1i) - e(1h,1i) - e(1h,0i)     \\
&E_{B,3}& = e(2h,1i) - e(0h,1i) - e(2h,0i)
\nonumber
\end{eqnarray}
\noindent
The physical
meaning of these quantities is quite obvious. $E_{B,2}$ is the usual
binding energy for the pure system.\cite{binding} $E_{B,3}$
corresponds to compare the state where the two holes are trapped to
the impurity with the state where the holes move away from the
impurity. $E^\prime_{B,2}$ is the energy difference between the state
with a hole pair away from the impurity and the state with one hole
bound to the impurity.
Thus, for the holes forming a bound pair not trapped by
the impurity the following inequalities should hold: $E_{B,2} < 0$,
$E^\prime_{B,2} < 0$, and $E_{B,3} > 0$.

The results for $E_{B,2}$, $E^\prime_{B,2}$ and $E_{B,3}$ obtained
for the $4 \times 4$ lattice and for the case of a nonmagnetic
impurity $J^\prime =0$ are shown in Fig. \ref{fig2}(a).
Very similar results are also obtained for the 18 sites tilted
cluster. It can be seen
that the conditions for the existence of a hole pair not bound to
the impurity are satisfied in the range $0.2 \leq J/t \leq 0.5$.
For a better understanding of the physical situation that emerges
from the study of $E_{B,2}$, $E^\prime_{B,2}$ and $E_{B,3}$, it is
important to examine also the hole-impurity binding energy $E_{B,1}$
as well as
\begin{eqnarray}
E_{B,4} = e(2h,2i) - 2 e(1h,1i)  .
\end{eqnarray}
\noindent
Notice that $E_{B,4}$ as $E^\prime_{B,2}$ reduces to
$E_{B,2}$ in the absence of impurities.
$E_{B,4}>0$ and $E_{B,1}<0$
means pair-breaking effect, each hole been trapped by
an impurity. Both quantities have been added to Fig. \ref{fig2}(a).
$E_{B,1}$ is negative for $J \geq 0.2$ implying that there is a
hole-impurity bound state. It is interesting to note that taking
into account the results for $E_{B,2}$, $E^\prime_{B,2}$ and $E_{B,3}$
discussed above this hole-impurity bound state only appears for
$J \geq 0.5$ when a second hole is added. However, as we shall show
below, the tendency to form a hole-impurity bound state is the main
source of suppression of pairing. Also interesting is the
result of $E_{B,4}$ indicating that two impurities split the hole
pair for $J \leq 0.3$.

The most important results for their connection to the experimental
results observed in the cuprates are shown in Fig. \ref{fig2}(b).
In this figure, we show $E^\prime_{B,2}$ and $E_{B,3}$ obtained for
the $4 \times 4$ cluster for several values of $J^\prime/J$. We also
add $E_{B,2}$ for comparison.
It can be seen that $E^\prime_{B,2}$
is suppressed as $J^\prime/J$ decreases from 1, corresponding to
the pure case except for the condition of exclusion of holes at the
impurity sites, to 0 which corresponds to the nonmagnetic case.
This behavior of the binding energy in the presence of impurities
is consistent with the experimentally observed one.
$E_{B,3}$ also decreases in such a way that the interval of $J/t$
where the bound state of holes exists and is not trapped to the
impurity narrows down as $J^\prime/J$ decreases.

Additional evidence of the strongest pair-breaker effect of
nonmagnetic impurities in the $t-J$ model comes from the study
of the quasiparticle weight of pairs and pairing correlations
with d$_{x^2-y^2}$  symmetry.
The quasiparticle weight of pairs $Z_{2h}$ is defined as
\begin{eqnarray}
Z_{2h} = \frac{|<\Psi_0^{2h} | \Delta |\Psi_0^{0h} > |}
{(<\Psi_0^{0h} | \Delta^\dagger \Delta |\Psi_0^{0h} > )^{1/2}}
\label{pairdyn}
\end{eqnarray}
\noindent
where $\Delta = \frac{1}{N} \sum_{i\ne i_0} \Delta_i$. The pairing
operator at site $i\ne i_0$ is
\begin{eqnarray}
\Delta_i = \sum_{\mu} g_\mu c_{i\uparrow} c_{i+\mu\downarrow},
\end{eqnarray}
\noindent
where the sum extends over the NN of site $i$ and
$g_\mu$ are the form factors that determine the pairing symmetry.
Note that when i corresponds to one of the four NN sites of $i_0$
the sum over $\mu$ is restricted to only $i+\mu\ne i_0$.
The rest of the notation is standard.\cite{pairdyn,prd}
The results obtained in the $4 \times 4$ lattice for d$_{x^2-y^2}$
symmetry as a function of $J/t$ and $J^\prime/J=0$ ,0.5 and 1.0.
are shown in Fig. \ref{fig3}(a). It can be seen that the largest
reduction of $Z_{2h}$ corresponds to the $J^\prime/J =0$ case.
Notice from the result for $J^\prime/J =0.5$ that this reduction is
not linear in $J^\prime/J$.
The pairing correlations are defined as
\begin{eqnarray}
C_{\Delta}(r) = \frac{1}{N} \sum_{i}
<\Delta_i^\dagger \Delta_{i+\bf r} >,
\label{paircorr}
\end{eqnarray}
\noindent
where, in the presence of an impurity, the sum is restricted to
$i$ and $i+r \ne i_0$.
It is well-known that close to half-filling the pairing correlations
at large distances are quite suppressed.\cite{dagorev} For this reason,
in Fig. \ref{fig3}(b) we show $C_{\Delta}(r)$ for $r \ge 1$ on the
$4 \times 4$ lattice at quarter filling with $J/t=3$ where they are
maximum in the pure system.\cite{dagrieprl}
The tail in $C_{\Delta}(r)$ is mostly suppressed for $J^\prime/J=0$.
The results for $J^\prime/J=0.5$ indicate again a nonlinear behavior
of pairing as a function of $J^\prime/J$.

In order to understand the origin of the pair breaking mechanism
let us examine the quantity $W_{p}$ defined as the probability
of finding the pair of holes in NN sites.
In Fig. \ref{fig4}(a) we show $W_{p}$ as a function of $J/t$ for
the pure system (full circles) and in the presence of a single
impurity with $J^\prime/J=0$ (full squares).
It is also shown for various $J^\prime/J$ the quantity $W_{p}$
computed with the constraint that the holes can not occupy the
NN sites of the impurity.
For the case of $J^\prime/J=0$, by comparing the total $W_{p}$
with the restricted one it can be concluded that there is a large
contribution to $W_{p}$ coming from the hole pair
bound to the impurity. This contribution becomes smaller as
$J^\prime/J$ is increased from zero.

To complete the characterization of the effects of impurities
on the ground state of the $t-J$ model we briefly discuss
their effect on the magnetic order\cite{sunmaki}
and on the dynamical magnetic fluctuations.
The dynamical spin structure factor at the AF wavevector $(\pi,\pi)$
shown in Fig. \ref{fig4}(b) exhibits, in comparison to the
pure case\cite{dagorev,yasu}, new structures at low energies.
Spectral weight is transfered from the peak at
$\omega\sim 0.8 J$, characteristic of the pure two-hole doped
$4\times 4$ system, to lower energies of order $J^\prime$.
This new resonance can be interpreted as the singlet-triplet
excitation energy of the singlet impurity-hole bound-state.
These results can be related to recent neutron scattering experiments
which have reported that the magnetic pseudogap at the AF wavevector
disappears with Zn doping.\cite{kakurai}

In summary, we have studied using exact diagonalization of the 2D
$t-J$ model in the presence of magnetic ($S=1/2$) and nonmagnetic
($S=0$) impurities. We have shown through an analysis of binding
energies, pairing correlations and dynamical pair susceptibility
that the nonmagnetic impurity has a stronger pair breaker effect than
the magnetic impurity. These results are in agreement with
experimental findings in Zn ($S=0$) and Ni ($S=1$) doped cuprates.
This agreement reafirms the ability of microscopic strongly
correlated electronic models to describe the physics of CuO$_2$
planes in the cuprates.
We have also given some indications of the possible essential role
of the binding of holes to impurities in suppressing superconductivity
in these systems.

D.P. thanks D.J. Scalapino and W. Hanke for many insightful
discussions. J.R. thanks E. Dagotto for useful comments.
Part of the numerical calculations were done at IDRIS (France).

%
%
\begin{figure}
\caption{
 $E_{B,1}$ vs $J^\prime/J$ for d-wave orbital symmetry,
for a spin 1/2 impurity (solid squares) and a spin 1 impurity
(open squares).
\label{fig1}}
\end{figure}

%
%
\begin{figure}
\caption{
a) $E_{B,2}$ (open squares), $E^\prime_{B,2}$ (open diamonds),
$E_{B,3}$ (solid circles), $E_{B,1}$ (solid squares) and
$E_{B,4}$ (solid triangles), defined in the text, as a function of
$J/t$ and for $J^\prime=0$.
b) $E_{B,3}$ (open symbols) and $E^\prime_{B,2}$ (solid symbols)
vs. $J/t$. The results correspond to $J^\prime=0$ (squares),
$J^\prime=J/3$ (circles), $J^\prime=2J/3$ (triangles) and $J^\prime=J$
(diamonds). We also show for comparison the result of $E_{B,2}$
(crosses).
\label{fig2}}
\end{figure}

%
%
\begin{figure}
\caption{
a) The quasiparticle spectral weight $Z_{2h}$ as a
function of $J/t$ for different values of $J^\prime/J$.
b) The pairing correlation $C_{\Delta}(r)$
at quarter filling for different values
of $J^\prime/J$. The results for the model with no impurities are
shown with solid circles.
\label{fig3}}
\end{figure}

%
%
\begin{figure}
\caption{
a) The probability $W_{p}$ vs. $J/t$. The results
for the pure system are shown with solid circles while
with solid squares we show the results with one
impurity in the system and $J^\prime=0$.
The results for $W_{p}$ with the exclusion of the NN
sites of the impurity are shown for various values of $J^\prime/J$
with open symbols as indicated on the plot.
b) Dynamical spin structure factor (arbitrary units) in the two
hole doped system for $J = 1.0$, and $J^\prime = 0$ (solid line),
0.25 (dashed line), 0.5 (dotted line) and 1.0 (thick solid line).
\label{fig4}}
\end{figure}

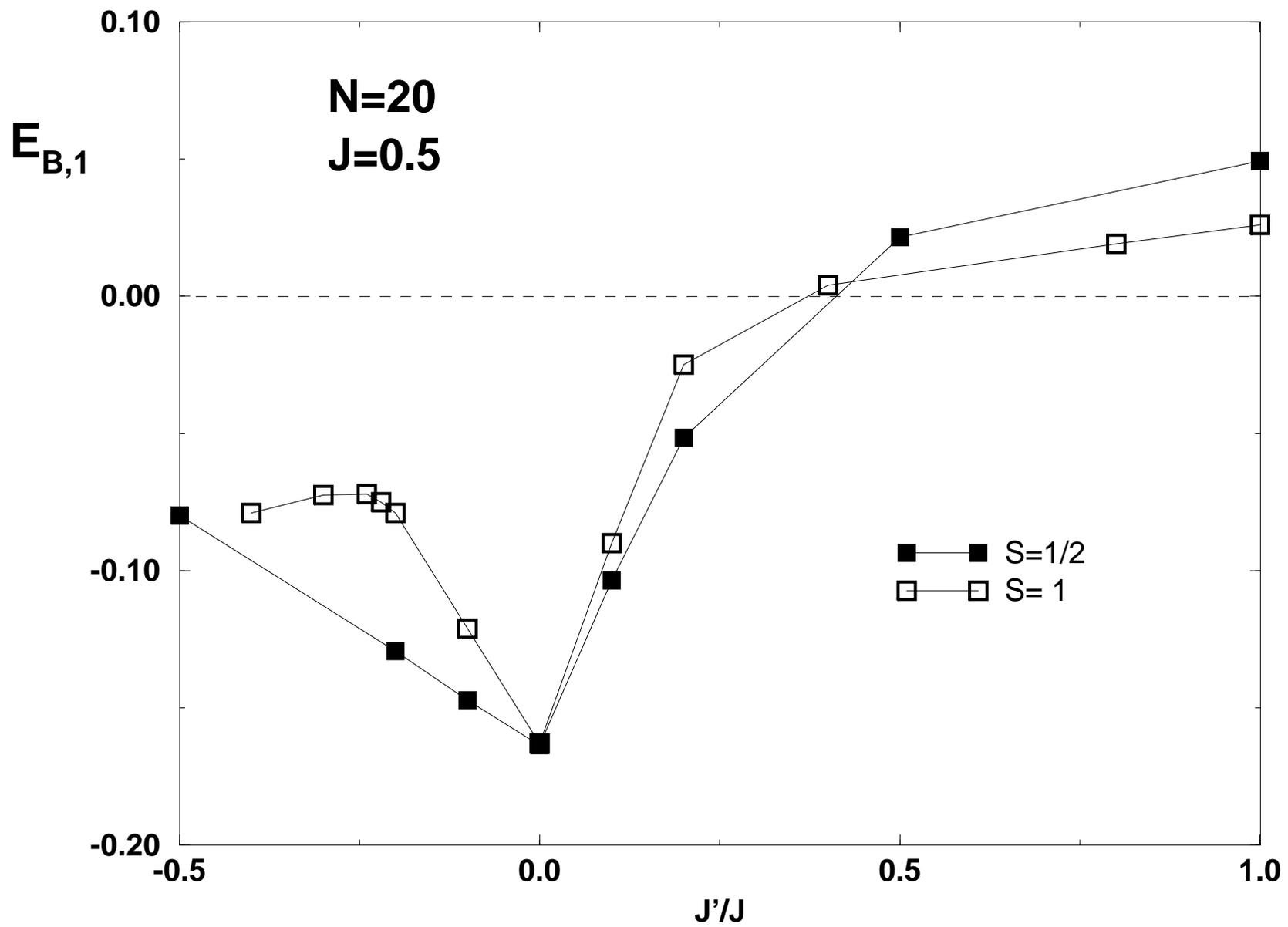
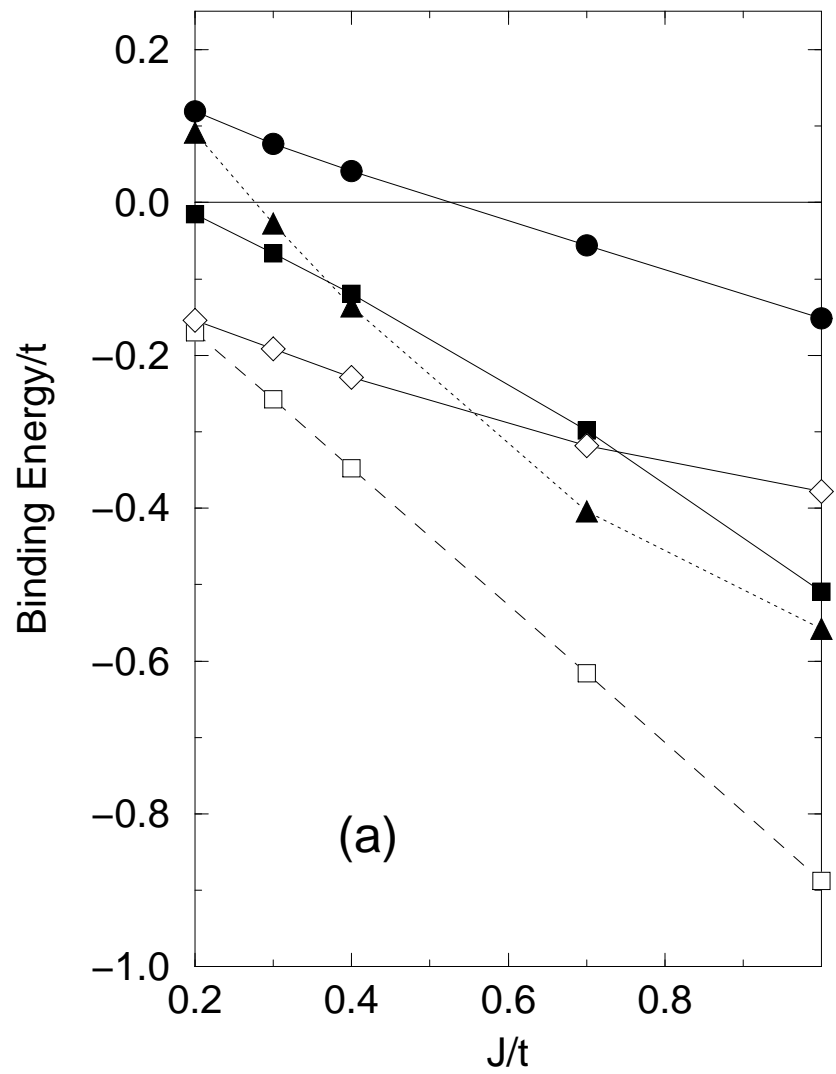
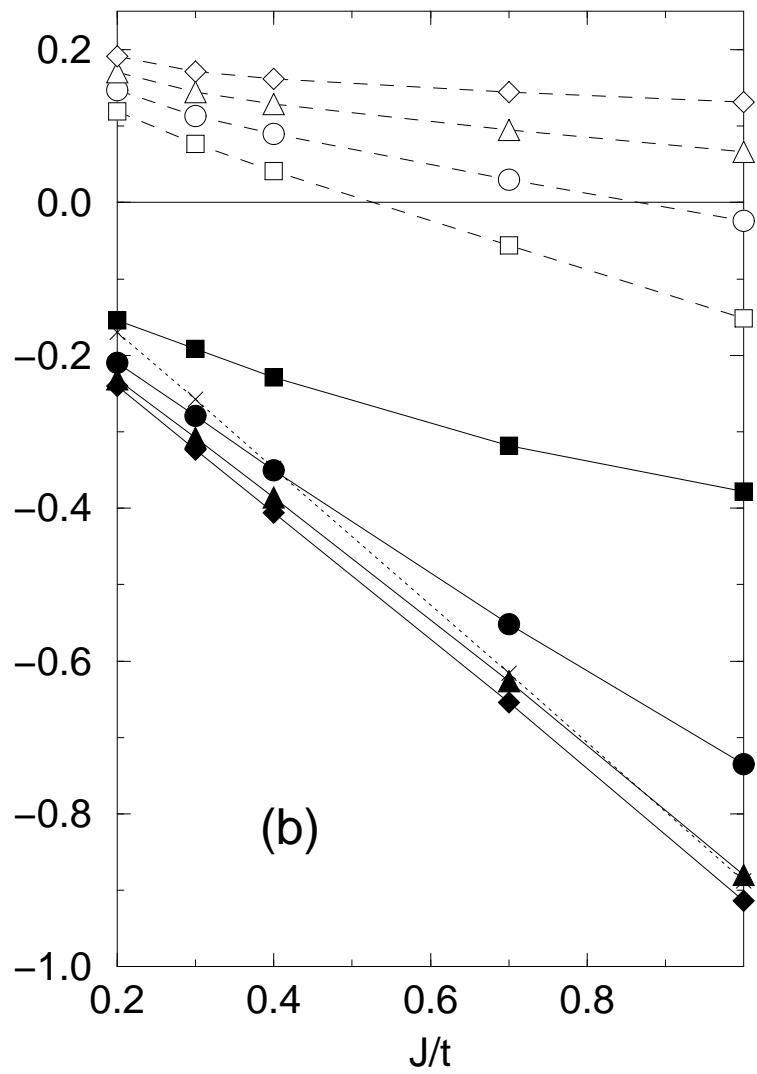
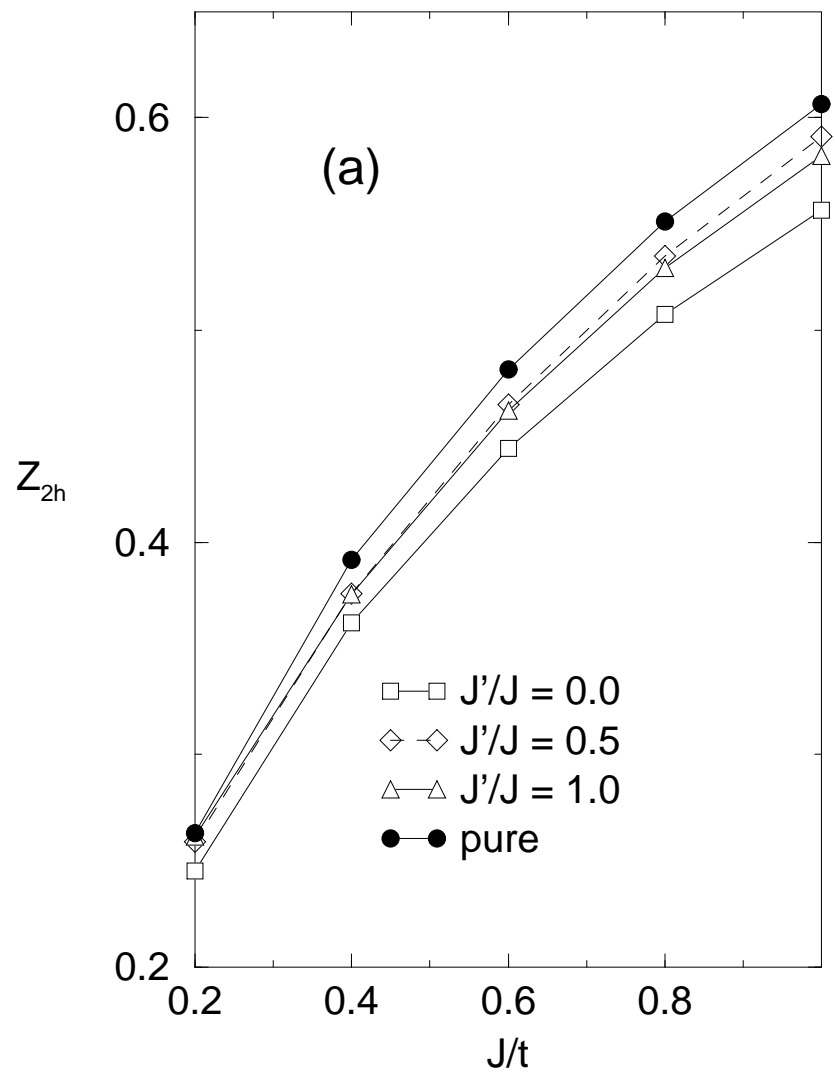
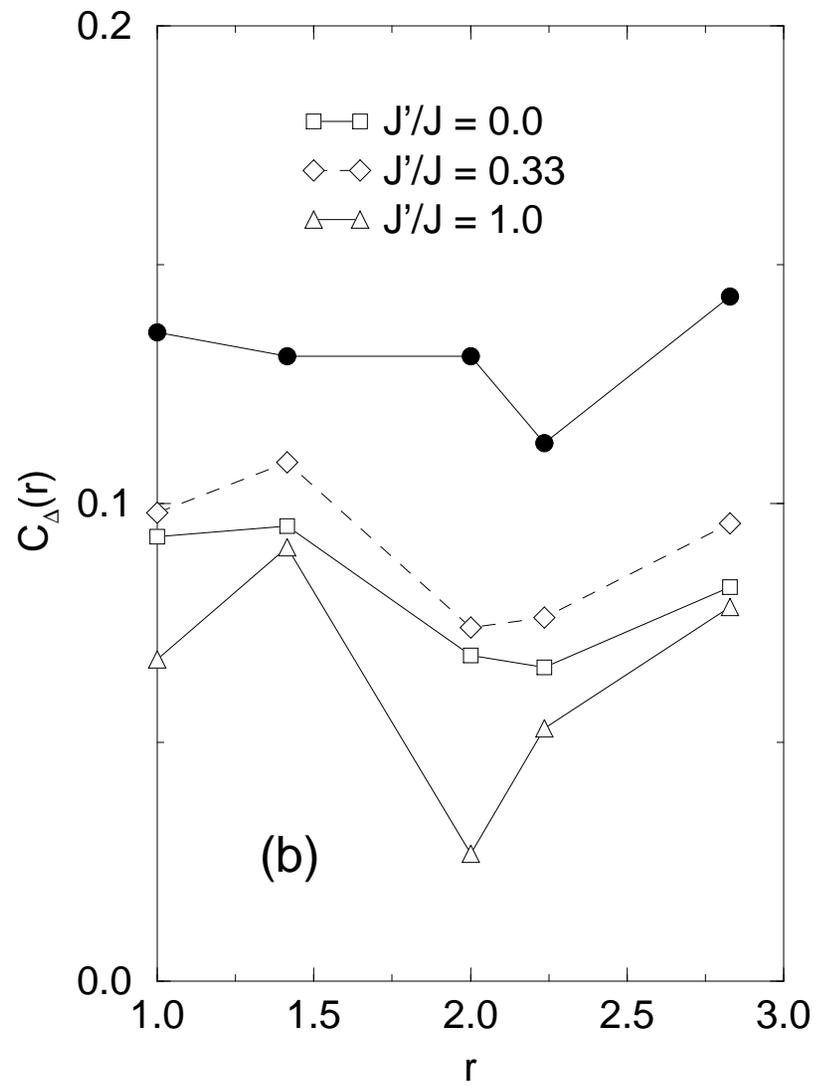
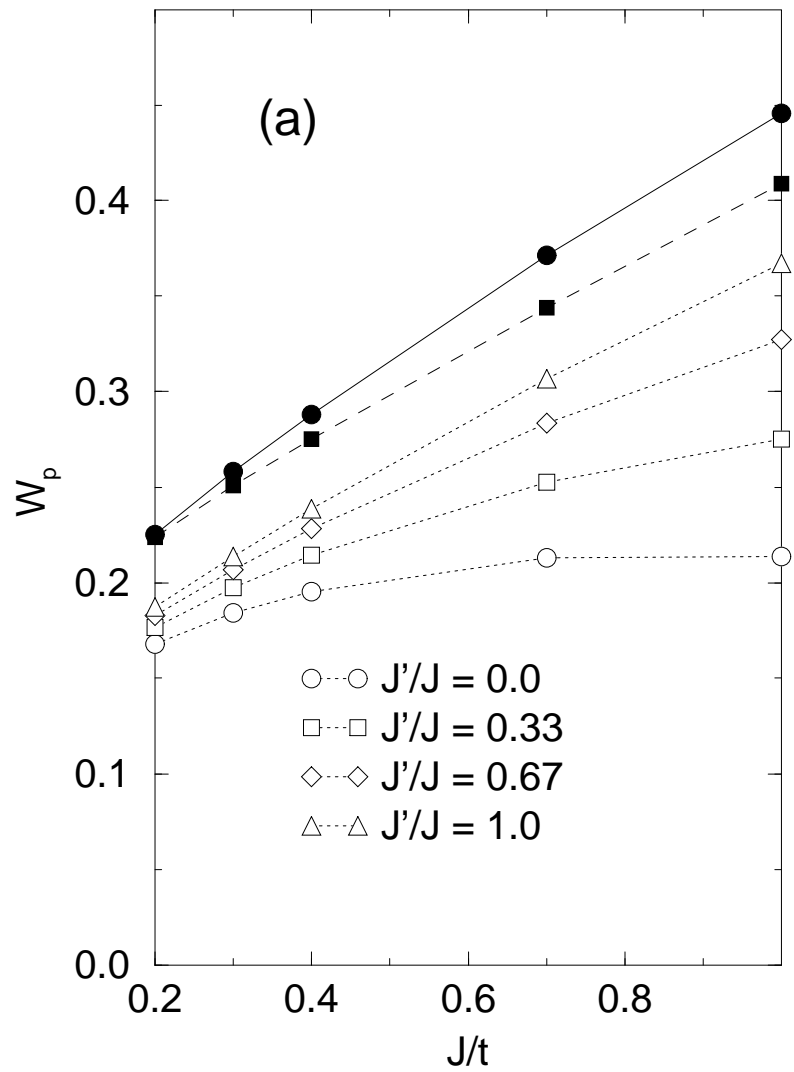
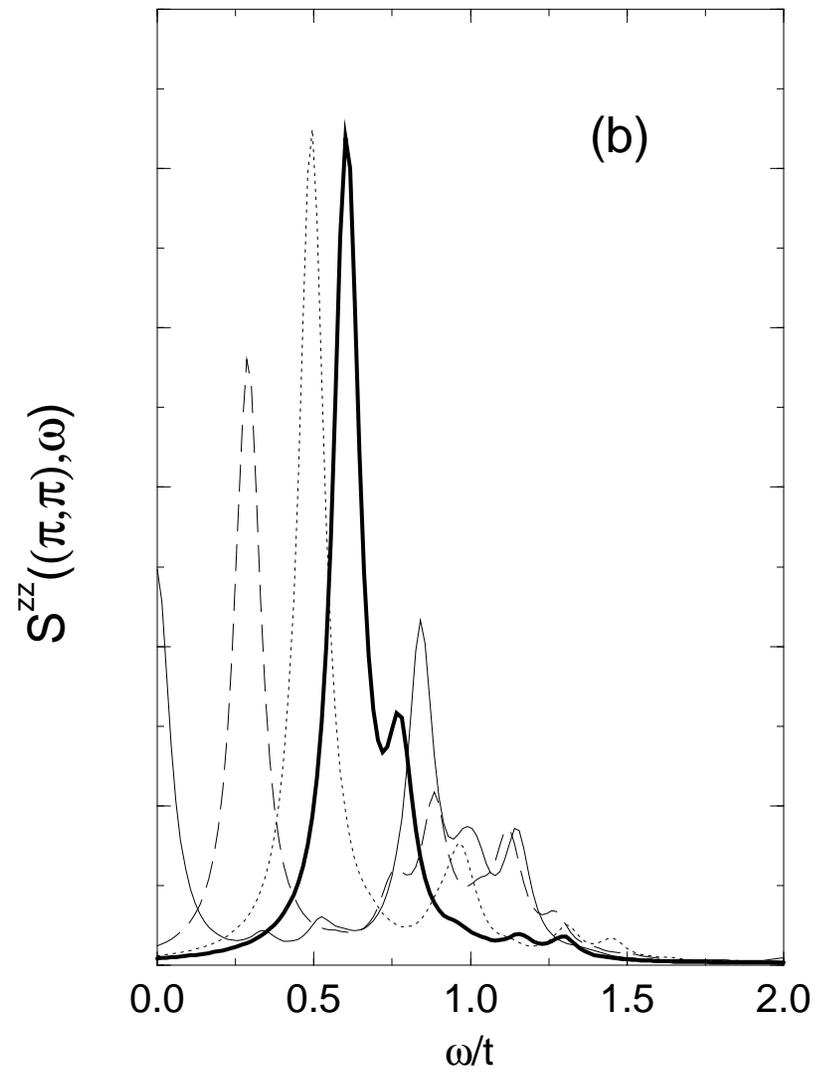
\end{document}